\documentclass{elsart}

 \usepackage{epsfig}

\usepackage{amssymb}

\newcommand{\be}{\begin{equation}}
\newcommand{\ee}{\end{equation}}
\newcommand{\bea}{\begin{eqnarray}}
\newcommand{\eea}{\end{eqnarray}}
\newcommand{\ba}{\begin{array}}
\newcommand{\ea}{\end{array}}

\begin{document}

\begin{frontmatter}



\hfill{NPAC-09-04}

\title{Neutrons and the New Standard Model}


\author{M.~J.~Ramsey-Musolf}

\address{Department of Physics, University of Wisconsin-Madison\\ Madison, WI 53711 USA\\ and \\ Kellogg Radiation Laboratory, California Institute of 
Technology\\ Pasadena, CA 91125 USA}

\begin{abstract}
Fundamental symmetry tests with neutrons can provide unique information about whatever will be the new Standard Model of fundamental interactions. I review two aspects of this possibility: searches for the permanent electric dipole moment of the neutron and its relation to the origin of baryonic matter, and precision studies of neutron decay that can probe new symmetries. I discuss the complementarity of these experiments with other low-energy precision tests and high energy collider searches for new physics. 
% Text of abstract
\end{abstract}

\end{frontmatter}

% main text
\section{Introduction}
\label{sec:intro}

The search for a unified framework for describing nature's fundamental interactions lies at the forefront of nuclear and particle physics. The Standard Model (SM) provides
a remarkably successful accounting for the strong and electroweak interactions, and it has survived numerous rigorous experimental tests at a variety of energy scales. The SM also 
contains many of the ingredients needed to explain a variety of astrophysical phenomena, such as weak interactions in stars, the rate of energy production in the sun, and possibly even the dynamics of supernovae. Nevertheless, it leaves a number of unanswered questions: Why is there more matter than antimatter in the universe? What makes up the elusive dark matter that constitutes roughly 25\% of the cosmic energy density? Was gravity unified with the other known forces at the birth of the universe and if so, how? Why is electric charge quantized? And why are the masses of neutrinos so tiny?

Answering these and other basic questions will require a \lq\lq new Standard Model" that leaves in tact its successes but puts them in a more comprehensive and fundamental context. This new SM will undoubtedly possess additional fundamental symmetries beyond those of the SM, such as supersymmetry or additional gauge symmetries -- both of which are suggested by constructions based on string theory. Uncovering these symmetries requires us to push the envelope at two frontiers: the high energy frontier, at which the Large Hadron Collider (LHC) is clearly aimed; and the high-precision frontier, for which low-energy studies such as those involving cold and ultracold neutrons will dominate the landscape for the next decade or more. 

In this talk, I will focus on the exciting prospects for discovery and insight at the precision frontier, with a particular emphasis on neutron physics. Given the limitations of space, I will consider in detail only two of the thrusts of the worldwide program: searches for the permanent electric dipole moment (EDM) of the neutron and studies of neutron decay. I will also illustrate the interplay of precision neutron studies with those being carried out using other low-energy precision tools, such as parity-violating electron scattering. More extensive discussions of fundamental symmetry tests with neutrons and their relation to other low- and high-energy experiments can be found in two recent reviews that I have coauthored\cite{Erler:2004cx,RamseyMusolf:2006vr}. Extensive references to the literature can be found there as well. 

\section{EDMs and the Origin of Matter}
\label{sec:edm}

Although the visible baryonic matter of the universe comprises only a small fraction of the total cosmic energy density, we have no definitive explanation of how this anthropically relevant form of matter came into being. It is generally assumed that the initial conditions for post-inflationary dynamics were matter-antimatter symmetric. If this assumption is valid, then the microphysics of the subsequently evolving cosmos would have to have produced the slight imbalance observed today. Characterizing this baryon asymmetry of the universe (BAU) as the baryon density to entropy density  $Y_B = n_B/s$, one has from measurements of the acoustic peaks in the cosmic microwave background\cite{ Dunkley:2008ie}
\begin{equation}
\label{eq:yb}
8.36 \times 10^{-11} 
< Y_B < 9.32 \times 10^{-11} (95\% \;
\textrm{C.L.})\ \ \ .
\end{equation}
This range is consistent with the value for $Y_B$ obtained from the light element abundances using Big Bang Nucleosynthesis\cite{Amsler:2008zz}. Forty years ago, Sakharov identified three ingredients needed in the early universe particle physics to account for this tiny number\cite{Sakharov:1967dj}: violation of baryon number (B); violation of both C and CP symmetries; and a departure from equilibrium dynamics in the cosmic thermal history as one might encounter in a phase transition associated with spontaneous symmetry-breaking. In principle, the SM contains all three ingredients. However, it is now known that the lower bound on the mass of the SM Higgs boson precludes any phase transition during the electroweak symmetry-breaking era\cite{Kajantie:1996qd,Kajantie:1996mn}. Even if such a transition were to have occurred, the effects of SM CP-violation are too strongly suppressed to allow for generation of particle asymmetry progenitors of the observed BAU. 

Clearly, elements of the new SM must make up for these SM shortcomings. If the see-saw mechanism for explaining the tiny scale of neutrino masses is correct, then the BAU might have been generated by out-of-equilibrium decays of the associated heavy right-handed Majorana neutrinos well before the electroweak symmetry-breaking era. This \lq\lq leptogenesis" scenario requires that neutrinos are, in fact, Majorana particles ($\nu={\bar\nu}$) -- thereby introducing lepton number violation --  and that the interactions of the early universe neutrinos violate CP invariance. Searches for neutrinoless double beta decay ($0\nu\beta\beta$)  test the possibility that neutrinos are their own antiparticles, while ongoing neutrino oscillation studies may uncover CP-violation in neutrino interactions. The results of these experiments will not be conclusive for leptogenesis, as there does not in general exist a direct relationship between the CP-violation in low-energy neutrino interactions and CP-violating interactions of neutrinos in the early universe. In short, leptogenesis is unlikely to be ruled out by laboratory studies, but its plausibility may be enhanced or diminished by these experiments.

A more testable possibility is the generation of the BAU during the electroweak symmetry-breaking era, a scenario known as electroweak baryogenesis (EWB). The coming decade is an exciting time for considering this possibility, as it requires new physics at the TeV scale that could be discovered in both LHC searches and low-energy precision tests. In particular, EWB requires a strong first order phase transition from unbroken to spontaneously broken electroweak symmetry, and new interactions involving scalar fields that facilitate this phase transition -- such as those appearing in extensions of the SM Higgs sector -- could be discovered at the LHC (see {\em e.g.}, Refs.~\cite{Profumo:2007wc,Quiros:1999jp} and references therein). Equally important is the possibility of new electroweak scale CP-violation that could be seen in searches for the permanent electric dipole moments of the neutron, electron, and neutral atoms\cite{Erler:2004cx,RamseyMusolf:2006vr,Pospelov:2005pr,Ellis:2008zy}. Both ingredients are critical, and there exists, therefore, a tight coupling between the LHC searches and EDM experiments in testing EWB. As I will illustrate shortly, the interpretation of EDM results in terms of the CP-violation needed for EWB will require input from the LHC, yet the absence of any signal for new physics at the LHC would not necessarily preclude new electroweak scale CP violation that could generate observable EDMs. 

The coupling between LHC and EDM probes of EWB can be illustrated by expressing the BAU and neutron EDM ($d_n$) in terms of the parameters of the new SM:
\begin{eqnarray}
\label{eq:yb1}
Y_B & = & \sum_k\, F_k(g_i, M_i; T, v_w, L_w,...)\, \sin\phi_k \\
\label{eq:dn1}
d_n & = & \sum_k\, H_k(g_i, M_i)\, \sin\phi_k\ \ \ ,
\end{eqnarray}
where $\phi_k$ represent all the complex CP-violating phases in the new SM, the $g_i$ and $M_i$ represent the couplings and masses in the theory; $T$ is the temperature at which electroweak symmetry breaking begins to take place (typically at around 100 GeV), and $v_w$ (\lq\lq wall velocity"), $L_w$ (\lq\lq wall thickness"), {\em etc.} are parameters that characterize the geometry and dynamics of the expanding regions of broken symmetry and that are implicitly functions of the parameters of the Lagrangian. Measurements provide us with experimental values for (limits on) $Y_B$ and $d_n$; the results of collider experiments and precision tests will provide values on the masses and couplings of the new SM that appear in the functions $F_k$ and $H_k$; and theory tells us just what these functions are. Testing the viability of EWB in a given scenario for the new SM then amounts to asking the following question: once we have collected the experimental information on $Y_B$, $d_n$, $g_i$, $M_i$, {\em etc.}, is there a set of CP-violating phases $\{\phi_k\}$ consistent with Eqs.~(\ref{eq:yb1}) and (\ref{eq:dn1}) ?

To illustrate the importance of knowing both the mass spectrum (from the LHC) and CP-violating phase, consider an example from supersymmetry. In the minimal supersymmetric Standard Model (MSSM), particle asymmetries are generated at the electroweak temperature by the scattering of \lq\lq superpartners" from the surface of regions of broken symmetry. These superpartner-wall interactions depend on the new CP-violating phases of the MSSM whose effects are not suppressed by the light quark Yukawa couplings as in SM CKM-matrix CP-violation. Additional interactions between the superpartners and the quarks and leptons present in the thermal bath convert these superpartner asymmetries into a net left-handed SM fermion density, $n_L$. Ultimately, it is topological transitions called sphaleron processes that turn $n_L$ into a non-vanishing baryon number density. For a given set of CP-violating interactions in the MSSM, the efficiency with which superpartner asymmetries convert into $n_L$ depends on the properties of the superpartners, such as their masses. 

This effect is illustrated below in Figure \ref{fig:yb}, where we show the value of $\sin\phi_\mu$ needed to produce the observed baryon asymmetry  as a function of the mass of the superpartner of the right-handed bottom quark. The phase $\phi_\mu$ is generally the most important phase for MSSM EWB in the simplest treatments of the vast MSSM parameter space. The plot is based on our recent computation\cite{Chung:2008ay}  of the function $F_\mu$ in Eq.~(\ref{eq:yb1}) that builds on our earlier treatment of the quantum transport during a supersymmetric electroweak phase transition\cite{Lee:2004we,Cirigliano:2006wh}. These results should be considered provisional, as there exists no treatment of the quantum transport that takes into account all of the relevant physical effects. The studies of Refs.~\cite{Carena:2000id,Konstandin:2005cd} have refined the computations of the CP-violating source terms in the transport equations, while our emphasis has been on computing the CP-conserving terms that are responsible for the transfer of particle number density from one species to another. At the end of the day, one would like a comprehensive computation of both types of terms that allows us to determine the MSSM-parameter dependence of all the terms in the transport equations, thereby providing a robust theoretical prediction for the function $F_\mu$. 

With this caveat in mind, one can see from Fig.~\ref{fig:yb} that as the right-handed bottom squark ${\tilde b}_R$ mass gets closer to that of the right handed top squark ${\tilde t}_R$, the value of $|\sin\phi_\mu|$ needed to generate the observed value of $Y_B$ diverges. This feature follows from a quenching of the baryon asymmetry in this region of parameter space as discussed in detail in Ref.~\cite{Chung:2008ay} . In the MSSM, one needs the mass of the ${\tilde t}_R$ to be $\mathcal{O}(100)$ GeV -- which we have assumed in this figure -- in order to produce the needed strong first order phase transition. Now consider a situation in which the LHC finds supersymmetry and determines that the masses of the other squarks, including those of the up- an down-quark superpartners, are relatively light. Then these results, together with limits on $d_n$, would create a conundrum for EWB in the MSSM. The lighter the squark masses, the larger the coefficient $H_\mu(g_i, M_i)$ of $\sin\phi_\mu$ in Eq.~(\ref{eq:dn1}) becomes, implying more stringent experimental limits on this phase. In fact, neutron and electron EDM limits yield stringent limits on $\sin\phi_\mu$ for light squarks and sleptons ($m_{\tilde f}\lesssim 1$ TeV): $|\sin\phi_\mu| \lesssim 0.01$ (95\% C.L.). On the other hand, if the masses of all the squarks are similar, then $m_{\tilde b_R}\sim m_{\tilde t_R}$, suppressing the function $F_\mu$ and requiring nearly maximal values of $\sin\phi_\mu$. 
Thus, we see that viability of EWB in the MSSM will depend on LHC discovery of a light ${\tilde t}_R$ and information implying that all the other squarks are heavy in order to evade the EDM limits while satisfying the requirements on $\sin\phi_\mu$. 

\begin{figure}[tb]
\begin{center}
\begin{tabular}{cc}
\includegraphics[scale=1,width=8.5cm]{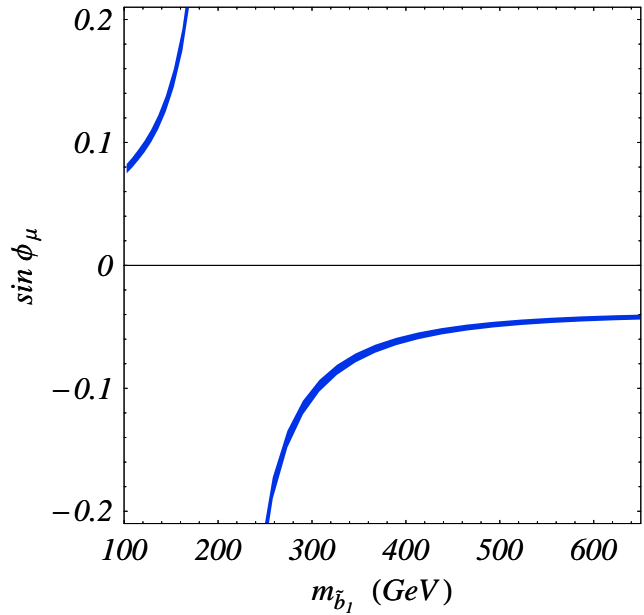}
\end{tabular}
\end{center}
\caption{Value of the CP-violating phase $\phi_\mu$ needed to obtain the WMAP value for the cosmic baryon asymmetry as a function of the right-handed bottom squark mass (labeled here $m_{\tilde b_1}$) for right handed stop mass of 150 GeV\cite{Chung:2008ay}. Figure courtesy of S. Tulin.}
\label{fig:yb}
\end{figure}

Naively, one might assume that one may evade present EDM bounds and any future null results for experiments performed with greater sensitivity by taking the superpartners to be sufficiently heavy. In practice, doing so will eventually preclude MSSM EWB since the superpartner of the Higgs bosons and electroweak gauge bosons (\lq\lq charginos" and \lq\lq neutralinos") must be relatively light in order to be sufficiently abundant at $T\sim 100$ GeV. In this case, one can encounter sizeable two-loop EDMs of the neutron and electron -- containing charginos and neutralinos and depending on $\phi_\mu$ --  even if the squarks and sleptons are much heavier than one TeV. Under the most optimistic theoretical scenario for MSSM EWB, one would need $|d_{e,n}|\gtrsim 10^{-28}$ $e$-cm in order to accommodate sufficiently light charginos and neutralinos and sufficiently large $\sin\phi_\mu$ to generate the experimentally observed BAU. EDMs of this magnitude could be observed in the neutron EDM experiments underway at ILL, PSI, and the SNS. Null results at this level of sensitivity would effectively rule out the minimal supersymmetric scenario for EWB. 

\begin{figure}[tb]
\begin{center}
\begin{tabular}{cc}
\includegraphics[scale=1,width=8.5cm]{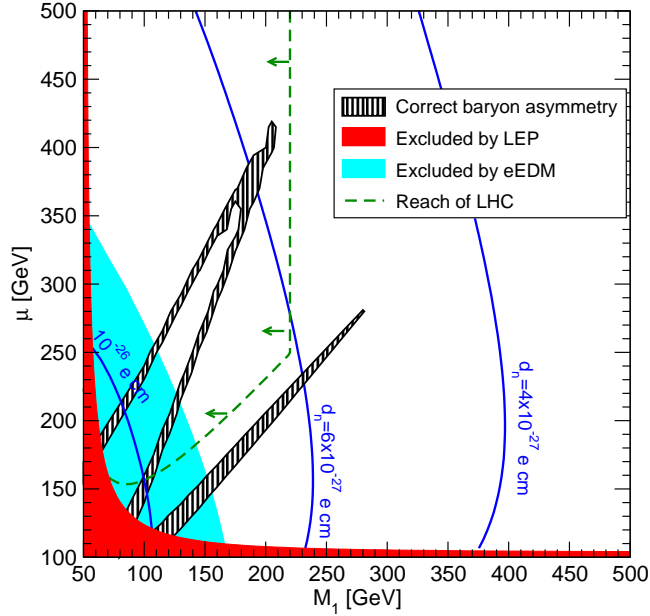}
\end{tabular}
\end{center}
\caption{Regions of the MSSM mass parameter space consistent with EWB for $\sin\phi_\mu=0.5$. Figure courtesy of S. Profumo.}
\label{fig:mu}
\end{figure}

I would like to emphasize two additional points on this topic. First, there exist scenarios under which MSSM EWB is driven primarily by the neutralinos and for which the LHC might be unable to discover the neutralinos because the associated missing energy is too small to be identified conclusively. This region of parameter space is nevertheless entirely accessible to the up-coming EDM searches, as illustrated in Fig.~\ref{fig:mu} (based on the work of Ref.~\cite{Cirigliano:2006dg}). There we show, for fixed $\sin\phi_\mu$, the regions in the space of MSSM mass parameters $\mu$ and $M_1$ that govern the character of the neutralinos. The red region has been excluded by LEP searches for charginos, while the light blue region is ruled out by existing limits on the electron EDM. The dark blue curves indicate the parameter space reach of a $d_n$ search at the level of sensitivity indicated. The parameter space that can be reached at the LHC, based on analyses performed to date, is to the left of the green dashed curve. Black hashed \lq\lq funnel-like" regions are those that would lead to the observed baryon asymmetry in the MSSM. The upper funnel corresponds to \lq\lq wino-driven" EWB that involves precursors of the charginos, while the lower funnel that lies outside the LHC reach represents \lq\lq bino-driven" EWB and involves exclusively precursors of the neutralinos. The latter scenario now appears to be the more viable of the two\cite{Li:2008ez}. The prospective neutron EDM experiments can probe both EWB-viable regions of this mass parameter space while the LHC appears to be unable to cover the pure neutralino region. Thus, the absence of a neutralino discovery at the LHC would not by itself preclude MSSM EWB, in which case it would still be up to the EDM searches to provide the definitive test. 

Second, it is entirely possible that EWB is driven by the dynamics of some other extension of the SM that has not yet been scrutinized theoretically or even invented. In some models, EWB is generated by particles that do not carry SM quantum numbers (so-called singlet extensions), and it is possible that these new particles are too difficult to identify at the LHC and that the CP-violating phases needed for baryon number production at finite temperature are too small at $T=0$ to generate observable EDMs (see, {\em e.g.}, Ref.~\cite{Kang:2004pp}). Although such scenarios are not generally favored, the results of LHC and EDM searches may force us to consider these speculative ideas -- along with leptogenesis -- more seriously.

\section{Precision Tests and Neutron Decay}
\label{sec:decay}

One advantage of the EDM searches is that EDMs are highly suppressed in the SM. The observation of a neutron EDM would be smoking gun evidence for either new CP-violation or a non-vanishing $\theta$-term in the QCD Lagrangian. The interpretation of precision measurements of SM-allowed processes, such as the SM-allowed correlations in neutron decay, requires a different perspective.  One requires that both the experimental error on a given observable $\mathcal{O}$, as well as the uncertainty in the theoretical SM prediction, be smaller than the expected size of effects from new physics. If this new physics occurs through loops, then one generally requires that
\begin{equation}
\label{eq: obs}
\frac{\delta\mathcal{O}}{\mathcal{O}^\mathrm{SM}} \lesssim \frac{\alpha}{\pi}\, \left(\frac{M}{{\tilde M}}\right)^2\ \ \ ,
\end{equation}
where $\delta\mathcal{O}$ is the combined experimental error and theoretical SM uncertainty, $M$ is the relevant SM mass scale, and ${\tilde M}$ is the mass of the heaviest new particle in the loop. For neutron decay, $M=M_W$, so that for any new particle with mass of order 100 GeV, one would need the combined relative experimental and theoretical uncertainty to be $\lesssim 10^{-3}$ in order to begin to discern deviations associated with the new physics. Although this requirement is quite stringent, the experimental neutron community has made impressive progress in pushing the level of precision to this level. It is, then, incumbent on theorists to achieve the corresponding level of confidence in our SM predictions and to quantify the possible effects of a variety of scenarios for the new SM.

The best-known illustration of these considerations involves tests of the unitarity of the CKM matrix involving its first row. This test requires the value of two derived quantities: $V_{ud}$ and $V_{us}$ (the third entry in the first row, $V_{ub}$, is too small to be relevant). The most precise value of $V_{ud}$ is obtained by comparing the vector coupling constant $G_V^\beta$ obtained from superallowed nuclear $\beta$-decay with the Fermi constant $G_\mu$ extracted from the muon lifetime:
\begin{equation}
\label{eq:vud}
\frac{G_V^\beta}{G_\mu} = V_{ud}\left(1+\Delta{\widehat r}_\beta^V-\Delta{\widehat r}_\mu\right) g_V(0)\ \ \ ,
\end{equation}
where $\Delta{\widehat r}_\beta^V$ contains the radiative corrections to the tree-level neutron decay amptlidue and $\Delta{\widehat r}_\mu$ contains the electroweak radiative corrections to the tree-level muon decay amplitude (QED corrections to the four-fermion interaction are explicitly included when extracting $G_\mu$ from the muon lifetime). The quantity $g_V(0)$ is the nucleon isovector charged current vector form factor at vanishing four momentum transfer squared. In writing Eq.~(\ref{eq:vud}) I have assumed that all many-body nuclear corrections have been taken into account so that the ratio corresponds to nucleon level physics. In principle, one should be able to compare the superallowed and neutron decay determinations of $G_V^\beta/G_\mu$.

The most precise value of $V_{us}$ is obtained from $K_{e3}$ decays: $K^+\to \pi^0 e^+\nu_e$. The partial rate for this decay $d\Gamma (K_{e3}) $ is proportional to the product of $|V_{us}|^2$ and the square of a kaon form factor, $f_+^K(0)$:
\begin{equation}
\label{eq:vus}
d\Gamma(K_{e3}) \propto | f_+^K(0)|^2 \, |V_{us}|^2\ \ \ ,
\end{equation}
where the other terms in the partial rate (not shown) include short- and long-distance electroweak radiative corrections, a phase space function, chiral SU(2)-breaking corrections, and a correction from the slope of the kaon form factor at the photon point. 

Obtaining $V_{ud}$ and $V_{us}$ from experimental measurements requires theoretical input as implied by Eqs.~(\ref{eq:vud},\ref{eq:vus}). The largest uncertainty in $V_{ud}$ is the theoretical uncertainty associated with computing the $W\gamma$  box graph contribution to $\Delta{\widehat r}_\beta^V$ that involves contributions from hadronic momentum scales that must be estimated or constrained. Recently, Marciano and Sirlin have reduced their estimate of the uncertainty in this contribution by a factor of two\cite{Marciano:2005ec} -- a substantial improvement. The situation with $V_{us}$ is more problematic, as the value of $f_+^K(0)$ must be determined theoretically. Two approaches have been used: computations using lattice QCD and analyses using chiral perturbation theory (ChPT) wherein the relevant unknown low energy constants have been obtained in the large $N_C$ (number of colors) limit. The approaches lead to different conclusions regarding whether or not the CKM unitarity is satisfied. If one uses the superallowed results for the left hand side of Eq.~(\ref{eq:vud}) one obtains\cite{RamseyMusolf:2006vr}
\begin{equation}
\label{eq:unitarity}
|V_{ud}|^2+|V_{us}|^2+|V_{ub}|^2=\cases{0.9968\pm 0.0014, & ChPT/large $N_C$ \cr
	0.9998\pm 0.0015, & lattice}
\end{equation}
where the two results correspond to the two approaches to obtaining $f_+^K(0)$ and where the contribution of $V_{ub}$ is negligible. 

The possibilities in Eq.~(\ref{eq:unitarity}) have several implications. Clearly, it is important for theory to provide a robust value of $f_+^K(0)$. After doing so, if unitarity is violated, then it is possible that there are either yet unknown effects in the SM radiative corrections to either $\beta$-decay or $K_{e3}$ decays or in the nuclear structure corrections that must be applied to obtain a value for $G_V^\beta$. Given the level of scrutiny the superallowed decays have received over many years\cite{Hardy:2005kr}, I consider the latter possibility unlikely. Nevertheless, various checks on the nuclear structure correction computations are being carried out through the study of  nuclear decays of even-even $I_Z=-1$ or odd-odd $I_Z=0$ nuclei\cite{Hardy:2005kr}. 
Ideally, the neutron decay experiments that exploit both the lifetime and one of the correlation parameters that depends on the axial to vector ratio $\lambda$ would provide an independent test, though the situation regarding the neutron lifetime needs to be resolved. 

At the end of the day, a unitarity deviation persists and no SM explanation is forthcoming, then the culprit could be new physics contributions to the difference $\Delta{\widehat r}_\beta^V-\Delta{\widehat r}_\mu$. In the context of supersymmetry, for example, loops involving superpartners could narrow the gap, though for this scenario to be realized the spectrum of superpartners would have to differ from expectations based on SUSY model building\cite{Kurylov:2001zx,RamseyMusolf:2006vr}. An alternate possibility is that the MSSM violates a discrete symmetry called \lq\lq R-parity", which amounts to a violation of $B-L$. The presence of R-parity violating (RPV) interactions allows for new tree-level superpartner exchange contributions to the $\beta$-decay and $\mu$-decay amplitudes that could readily account for any unitarity deviation without requiring supersymmetric and a non-standard superpartner spectrum\cite{RamseyMusolf:2000qn,RamseyMusolf:2006vr}. The price for this solution is rather high: the lightest superpartner would not be stable, and we would have to give up supersymmetric dark matter. The violation of this symmetry would also imply that the neutrino is a Majorana particle, implying a non-standard interpretation of $0\nu\beta\beta$ searches.  Either way, the implications would be quite interesting and illustrate the importance of achieving a robust test of CKM unitarity.

Beyond this test, studies of the neutron decay correlation coefficients can provide a unique window on the possible symmetries of the new SM. In the MSSM, for example, correlations that depend on the ratio of electron mass and energy, such as the Fierz interference coefficient or the $\beta$ energy-dependent part of the neutrino asymmetry parameter \lq\lq $B$" (not to be confused with baryon number) are particularly interesting. These non-$(V-A)\otimes (V-A)$ effects are generated at the one-loop level through box graphs, and they are distinct from the radiative corrections in the MSSM that retain the purely left-handed structure of the SM charged current interaction and could affect the difference $\Delta{\widehat r}_\beta^V-\Delta{\widehat r}_\mu$. 

It is useful to characterize non-$(V-A)\otimes (V-A)$ interactions using the effective low-energy Lagrangian 
\begin{equation}
\mathcal{L}=-\frac{4 G_\mu}{\sqrt{2}}\, \sum_{\gamma,\epsilon,\delta} a^\gamma_{\epsilon\delta}\, {\bar e}_\epsilon\gamma^\gamma \nu {\bar u}\gamma_\delta d_\delta\ \ \ ,
\end{equation}
where the sum runs over all independent Dirac matrices ($\gamma$) and chiralities of the electron ($\epsilon$) and d-quark ($\delta$). The supersymmetric box graphs can generate non-vanishing contributions to the scalar interactions having strengths $a^S_{RR}$ and $a^S_{RL}$ and the tensor interaction with coefficient $a^T_{RL}$\cite{Profumo:2006yu}. The new scalar interactions, in turn, induce a non-vanishing Fierz interference coefficient
\begin{equation}
b_F^{\rm SUSY\,  box} = \pm \frac{2 g_S}{g_V}\left(\frac{a^S_{RL}+a^S_{RR}}{a^V_{LL}}\right)\ \ \ ,
\end{equation}
where $a^V_{LL}=V_{ud}$ at tree-level in the SM; $g_S$ ($g_V$) are the nucleon charged current scalar (vector) form factors; and the upper (lower) sign corresponds to $\beta^-$ ($\beta^+$) decay. Similarly, one obtains an energy-dependent term in the neutrino asymmetry parameter
\begin{eqnarray}
B^{\rm SUSY\,  box} & = & -2\left(\frac{\Gamma m_e}{E}\right)\, \frac{\lambda}{1+3\lambda^2}\,
{\rm Re}\, \Biggl\{ 4\lambda \left(\frac{g_T}{g_A}\right)\, \left(\frac{a^{T}_{RL}}{a^{V}_{LL}}\right)^\ast\\
\nonumber
&& +\left[2 \left(\frac{g_T}{g_A}\right)\,  \left(\frac{a^{T}_{RL}}{a^{V}_{LL}}\right)^\ast - \left(\frac{g_S}{g_V}\right)\, 
\left(\frac{a^{S}_{RL}+a^S_{RR}}{a^{V}_{LL}}\right)^\ast\right]\Biggr\}
\end{eqnarray}
where $g_T$ is the nucleon tensor form factor, $\Gamma=\sqrt{1-(Z\alpha)^2} =1$ for neutron decay, and $E$ is the $\beta$ energy. 

When the superpartner masses are of order of the electroweak scale and the mixing between the left- and right-handed scalar superpartners of the first generation fermions is nearly maximal, the SUSY box-graph contributions to $b_F$ and the coefficient of the $m_e/E$ term in the neutrino asymmetry parameter can be of order $10^{-3}$. SUSY model builders and phenomenologists normally assume that the scalar fermion mixing is considerably smaller, arising from the same Yukawa couplings that generate the fermion masses. This assumption has not been tested experimentally for the first generation sfermions. If this mixing were observed experimentally it would require that all but the SM-like SUSY Higgs boson be super heavy in order to avoid a vacuum that is not color neutral\cite{Profumo:2006yu}. Measurements of these energy-dependent correlation coefficients with sensitivity approaching the $10^{-4}$ level could either discover large mixing effects or provide experimental confirmation for theoretical assumptions that go into SUSY model building and related phenomenology (for a discussion of neutron decay probes of other new physics scenarios, see {\em e.g.}, Ref.~\cite{Herczeg:2001vk}). 

\section{ Neutron Decay and other Fundamental Symmetry Tests}
\label{sec:other}

It is clear that fundamental symmetry tests with neutrons can provide information that is both interesting and complementary to high energy searches for the new SM. What about their relation with other low-energy, high precision tests of fundamental symmetries? From the standpoint of CP-violation, the neutron EDM searches are essential. If, for example, a non-zero result is obtained for an atomic EDM, we would not know whether the source is the chromo-EDM of the quark that generates long-range pion-exchange contributions to the EDMs of hadronic systems, the QCD $\theta$-term, or a new CP-violating electron-quark interaction. The presence or absence of a non-vanishing neutron EDM would narrow down the set of possibilities. Similarly, the observation of a non-vanishing electron EDM would likely indicate the presence of new CP-violation in the extended electroweak sector of the SM, but would not exclude non-vanishing chromo-EDMs or $\theta$-term effects. The full complement of EDM searches will be needed either to identify the type of previously unseen CP-violation or exclude most new CP-violating possibilities associated with electroweak scale new physics, at least if the latter is responsible for the origin of visible matter.

The interpretation of neutron decay experiments have a similar complementarity with other low-energy probes. For example, if a CKM unitarity violation is conclusively established, and if one postulates that new RPV interactions in SUSY are responsible, then one would also expect to see deviations from SM predictions for the parity-violating asymmetries in polarized elastic electron-proton, M\o ller scattering, or deep-inelastic electron-neutron scattering. The presence of R-parity violation would also immediately imply that neutrinos are Majorana fermions and could alter  the conventional interpretation of $0\nu\beta\beta$ in terms of the absolute mass scale of the light neutrinos. Large $\beta$ energy-dependent effects in the Fierz interference or neutrino asymmetry parameter could complicate life for SUSY enthusiasts or point to the need for additional new symmetries beyond those of mainstream supersymmetry.

In short, fundamental symmetry tests with neutrons have tremendous potential to provide unique contributions to our knowledge of the new SM. The name of the game is precision: pushing the neutron EDM sensitivity to $\sim 10^{-28}$ $e$-cm and neutron decay studies to the few $\times 10^{-4}$ level is the next frontier. Given both the growth of activity in this field and the presence of high caliber physicists, there is no doubt in my mind that this frontier can be reached. 

\noindent {\bf Acknowledgments} I would like to thank the generosity of the Institute Laue-Langevin for partial support of my participation in the workshop and my colleagues for many enlightening conversations. I also thank my collaborators S. Profumo and S. Tulin for providing the figures. This work was supported in party by U.S. Department of Energy Contract DE-FG02-08ER41531 and by the Wisconsin Alumni Research Foundation. 

% \appendix

% \section{}
% \label{}

\end{document}